\documentclass[3p,twocolumn,sort,compress]{elsarticle}

\usepackage[T1]{fontenc}
\usepackage{lmodern}
\usepackage[normalem]{ulem}

\usepackage{graphicx}
\usepackage{amsmath}
\usepackage{amssymb}
\usepackage[x11names,dvipsnames]{xcolor}  
\usepackage[colorlinks=true,citebordercolor=teal,allcolors=teal]{hyperref}
\usepackage{mathpazo}  %
\usepackage{microtype}
\usepackage{siunitx}
\usepackage{xfrac}
\begin{document}

\begin{frontmatter}

\title{
Revisiting quantum effects on dislocation glide in bcc metals from DFT calculations and machine-learning potentials
}

\author[label1]{Arnaud Allera\corref{cor1}}
\ead{arnaud.allera@asnr.fr}
\author[label3]{Lisa Ventelon}
\ead{lisa.ventelon@cea.fr}
\author[label3]{Mihai-Cosmin Marinica}
\ead{mihai-cosmin.marinica@cea.fr}
\author[label2]{David Rodney}
\ead{david.rodney@univ-lyon1.fr}
\author[label3]{Laurent Proville}
\ead{laurent.proville@cea.fr}

\address[label1]{ASNR/PSN-RES/SEMIA/LSMA Centre d'études de Cadarache, F-13115 Saint Paul-lez-Durance, France}
\address[label3]{Université Paris-Saclay, CEA, Service de recherche en Corrosion et Comportement des Matériaux, SRMP, 91191 Gif Sur Yvette, France}
\address[label2]{Univ. Lyon, UCBL, Institut Lumière Matière, UMR CNRS 5306, F-69622 Villeurbanne, France}
\cortext[cor1]{Corresponding author}

\begin{abstract}
\noindent 
Quantum zero-point effects have been long proposed to explain a well-known discrepancy between the low-temperature flow stresses of body-centered cubic metals and corresponding atomistic models of plastic flow. Previous investigations on quantum effects relied on empirical interatomic potentials, which poorly reproduce dislocation energy landscapes compared to density functional theory (DFT) calculations.
Here, we revisit this problem using DFT and machine-learning interatomic potentials (MLIPs).
We show that while quantum effects do contribute to dislocation glide at low temperature, their magnitude is much lower than previously reported, and insufficient to reconcile atomistic predictions with experiments. 
Our results thus reopen a long-standing question and challenge for predictive atomistic modeling, on a fundamental property of crystals.

\end{abstract}

\end{frontmatter}

\section{Introduction}

At low temperature, the plastic yield of body-centered cubic (bcc) metals such as iron, tungsten, or tantalum is controlled by the resistance of the crystal lattice to the motion of screw dislocations.
To overcome this resistance, dislocations with a $\sfrac{1}{2}\,\langle111\rangle$ Burgers vector glide through the thermally activated nucleation and propagation of kink-pairs, allowing them to 
jump from a low energy position\textemdash a Peierls valley\textemdash to the next \cite{guyot1967critical,caillard2003thermally,seeger2022peierls}. 
A key quantity describing this process across multiscale models \cite{cereceda2016unraveling,bienvenu2022ab,ZHAO2024105640} is the Peierls stress $\tau_P$, which corresponds to the stress required to move a dislocation without the assistance of thermal fluctuations \cite{kubamoto1979thermally,brunner2000plastic,Caillard2010}.
The Peierls stress can be extrapolated from cryogenic deformation tests \cite{kubamoto1979thermally,kamimura2013experimental,Altshuler1967,glen1956xxxix,aono1981plastic,Suzuki1970Effect}, or computed directly by atomistic simulations, using either \SI{0}{\kelvin} molecular dynamics simulations or minimum enthalpy path calculations (see for instance \cite{rodney2007activation,rodney2017ab,allera2025activation,clouet2021screw} and Refs. therein).
It corresponds to the stress at which the activation enthalpy for kink-pairs vanishes.
However, despite decades of effort,
atomistic simulations consistently fail to quantitatively reproduce this fundamental property of bcc metals.
Whether based on empirical potentials \cite{chaussidon2006glide,proville2012quantum}, density functional theory (DFT) \cite{ventelon2013ab,weinberger2013peierls,dezerald2016plastic}, or more recently machine-learning interatomic potentials (MLIPs)  trained on DFT \cite{maresca2018screw,allera2025activation}, simulations systematically overestimate $\tau_P$ compared to experiments, by a factor of 2-3 \cite{groger2007explanation}. 
This discrepancy has been known since the earliest dislocation studies \cite{basinski1971,Suzuki1970Effect}, and is sometimes discarded by rescaling atomistic estimates using an arbitrary factor \cite{tallman2019hierarchical}.

Quantum effects were long postulated to ease dislocation motion at low temperature, either through quantum tunneling \cite{kubamoto1979thermally,glen1956xxxix,Mott1956-MOTLCI} or zero-point fluctuations \cite{Suzuki1970Effect,natsik1972influence,alefeld1964rate}, which are not included in classical atomistic calculations even at 0~K. 
A numerical assessment of quantum effects on dislocation glide, initially only accessible through simplified interaction potentials \cite{Suzuki1970Effect}, became possible with the introduction of empirical potentials that faithfully reproduce the dislocation core structure and Peierls barrier of screw dislocations.  
Based on such empirical potentials, some of the present authors argued that quantum zero-point fluctuations could drastically reduce the activation enthalpy \cite{proville2012quantum}. While other effects are known to contribute \cite{groger2007explanation,kamimura2018peierls,kang2012singular}, this quantum effect has since been largely accepted by the community as a likely explanation for the discrepancy between atomistic calculations and experimental estimates of the Peierls stress in bcc metals \cite{weygand2015multiscale,caillard2018geometry}.
Within the 0 K limit temperature, it was shown that zero-point fluctuations reduce the enthalpy barrier for kink-pair
activation by:
\begin{equation}
\delta E(\tau) = \frac{h}{2} \left ( \sum_i \nu_i^{\rm init} - \sum_j \nu_j^{\rm saddle} \right ),
\label{eq:DE}
\end{equation}
where $h$ stands for Planck's constant and where $\nu_i^{\rm init}$ and $\nu_j^{\rm saddle}$ are the vibration frequencies 
of a crystal containing either a straight dislocation at the bottom of a Peierls valley, or
a dislocation forming a single critical kink-pair, respectively.
Eq. \ref{eq:DE} corresponds merely to the zero-point energy difference between the initial and saddle states of the dislocation.
We emphasize that the present work concerns the thermally activated motion of dislocations below the Peierls stress, where the quantum correction originates from the change in crystal vibrations as the dislocation transits between adjacent Peierls valleys. This is to be distinguished  from  dislocations interaction with phonon wind that occurs above the Peierls stress when the dislocation is gliding continuously with a drag force inherent to phonon scattering
\cite{li2017nonperturbative,lerma2025quantum,Blaschke2021}. 
Using empirical embedded atom method (EAM) potentials, it has been shown quantitatively in Ref. \cite{proville2012quantum} that including $\delta E$ within harmonic transition state theory (TST) can substantially lower the predicted Peierls stress in bcc Fe, thereby reconciling simulations with experiments. In some cases, it was even found that the corrected Peierls stress\textemdash referred to as the quantum Peierls stress\textemdash could fall below experimental estimates when combined with non-Schmid effects \cite{barvinschi2014quantum}.

However, this appealing picture rests on a fragile foundation: the underlying interatomic potentials. While EAM potentials have allowed large-scale dislocation simulations that have been instrumental in advancing our understanding of dislocation physics, they are known to have limited quantitative accuracy for dislocation core properties and can even lead to qualitative artifacts. 
A notable example is the structure of the screw dislocation core in pure bcc metals, which is not asymmetrically spread as predicted by early potentials, but is instead compact and symmetric according to \textit{ab initio} calculations \cite{ventelon2007core,ventelon2013ab,rodney2017ab}.
It also appears that EAM potentials generally predict energy landscapes that are more rugged than DFT or MLIPs, leading to exaggerated anharmonic effects. In particular, we have recently shown that EAM potentials predict a likely unphysical stress dependence of the activation entropy $\Delta S$  and a strong anharmonicity even at low temperature \cite{allera2025activation}. Given the close connection between entropic (see Ref. \cite{allera2025activation}) and quantum contributions that are both related to vibrational frequencies of the crystal, this raises a critical question: are the large quantum corrections reported in earlier studies a genuine physical effect or an artifact of EAM potentials?

Until recently, this question could not be addressed directly because DFT calculations were too computationally demanding to allow computing the Hessian matrices required to determine the crystal vibrational mode frequencies. This limitation is now being lifted by advances in computational capabilities and through the advent of MLIPs\cite{goryaeva2021efficient,allera2025activation}. Taking advantage of such progress, we revisit the quantum effects on dislocation glide using state-of-the-art DFT, combined with
MLIPs interpolating a large set of DFT calculations presented in Ref.~\cite{allera2025activation}. 
Our results challenge the prevailing view, as we show that the zero-point energy correction in Fe and W is smaller than previously reported and can only account for a fraction of the discrepancy between simulations and experiments. Rather than resolving the problem, quantum effects appear to have been overstated, calling for a reassessment of their role in the plasticity of bcc metals.

\section{Results}

\subsection{DFT calculations on straight dislocations}

\begin{figure*}[htb!]
    \centering
\includegraphics[width=0.28\linewidth]{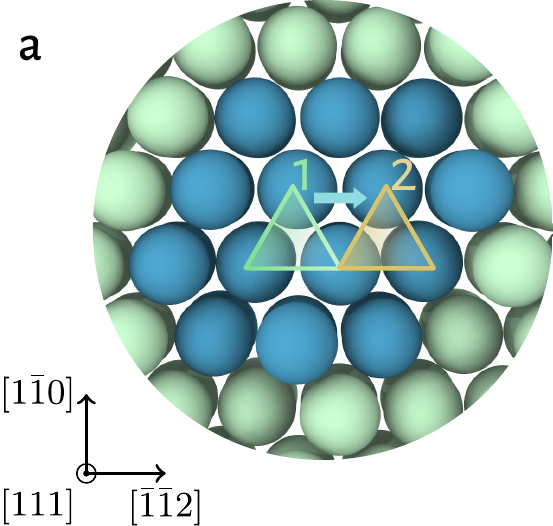} \hspace{0.25cm}
 \includegraphics[width=0.58\linewidth]{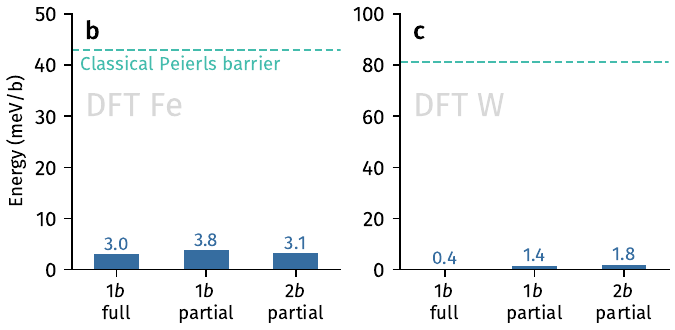}
    \caption{\textbf{Quantum correction to the Peierls barrier computed with DFT in Fe and W.}
(a) Atomic structure in the vicinity of the dislocation core. Core positions in the initial (1) and final (2) states are indicated by triangles.
(b–c) Quantum corrections per unit length in Fe and W for dislocation segments of length $1b$ and $2b$. “Full” denotes calculations in which all atoms are displaced to construct the Hessian, whereas  with “partial” only 14 atoms per Burgers vector in the vicinity of the dislocation core (highlighted in blue in (a)) are displaced {(see Supplementary Materials).}
}
\label{fig1}
\end{figure*}

We start by evaluating, in both Fe and W, the zero-point quantum correction to the Peierls barrier, i.e. to the energy barrier that a perfectly straight dislocation must overcome in order to glide between two successive equilibrium positions. Owing to translational invariance along the dislocation line, the simulation cell can be reduced to one Burgers vector ($1b$) in this direction, making DFT calculations tractable. Although this approach does not account for the kink-pair mechanism governing screw dislocation glide at finite temperature, it represents a useful reference accessible to DFT for assessing the magnitude of the quantum correction relative to the energy barrier.

In order to satisfy periodic boundary conditions used in DFT,
we consider a periodic array of dislocation dipoles arranged in a quadrupolar geometry, using simulation cells containing 135 atoms per Burgers vector along the dislocation line \cite{rodney2017ab,clouet2021screw}. Both dislocations are placed in  successive Peierls valleys in the initial and final configurations. The transition path is computed using the VASP code~\cite{kresse1996efficiency} to determine the energy of the system and the forces on each atom,
within a climbing nudged elastic band (NEB) simulation \cite{henkelman2000, henkelman2000climbing} (see Methods). 
Applying finite displacements in the 3 directions of space on each atom, the Hessian matrix is determined and subsequently diagonalized to obtain the frequencies of the vibrational modes.
The reader is referred to the Methods section for further details about these calculations.

\begin{figure*}[htb!]
    \centering
    \includegraphics[width=0.95 \linewidth]{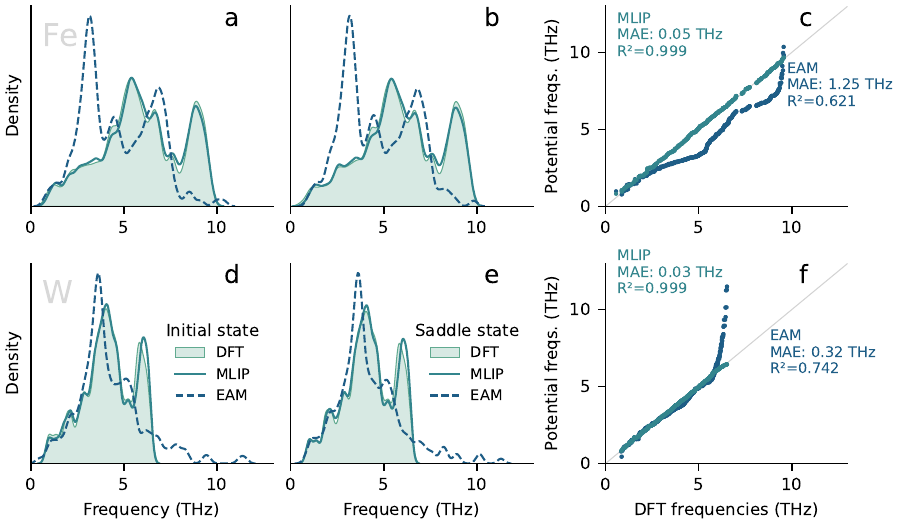}
    \caption{\textbf{Comparison of vibrational densities of states from DFT, MLIPs, and EAM potentials}
Vibrational densities of states for the initial (a, d) and saddle (b, e) configurations are shown for Fe (a–c) and W (d–f), computed using DFT, MLIPs, and EAM potentials. Panels (c) and (f) present a mode-by-mode comparison of frequencies obtained with the interatomic potentials against DFT. 
    }
    \label{fig2}
\end{figure*}

With a periodic segment of length one Burgers vector along the $\langle 111 \rangle$ direction, all atoms belonging to a given $\langle 111 \rangle$ column are constrained by periodicity to move in phase, severely restricting the accessible vibrational modes. To mitigate this artifact, we repeated the calculations using a dislocation of length two Burgers vectors ($2b$). In this case, atomic displacements along a $\langle 111 \rangle$ column are no longer constrained to be synchronized.
However, while the DFT calculation of the full Hessian matrix is feasible for a $1b$ dislocation, it becomes prohibitively expensive in the $2b$ cell. For this reason, in the latter case, we computed a partial Hessian in which only the atoms in the vicinity of one of the dislocation cores were displaced~\cite{barvinschi2014quantum}. 
{We verified that including only one layer of atoms around the core (shown in blue in Fig.~\ref{fig1}(a)) led to sufficient accuracy (see the Supplementary Materials for details).}
The full Hessian was therefore computed with DFT only for the $1b$ dislocation, while partial Hessians were computed for both $1b$ (for comparison) and $2b$. 
We note that in full calculations, the initial and saddle configurations are equivalent for both dislocations in the cell, and the quantum correction was divided by a factor 2, which is not needed for partial Hessian. 

The quantum zero-point energy variation along the transition path is reported for Fe in Fig.~\ref{fig1}(b) and for W in Fig.~\ref{fig1}(c) in the \SI{0}{\kelvin} limit, given by Eq.~(\ref{eq:DE}) in the absence of applied stress, considering full or partial Hessian matrices. 
For reference, we also indicate the classical Peierls barriers as dashed lines.

Comparison between the full and partial calculations in the  $1b$ simulation cell shows that, although the partial calculation deviates somewhat from the full one—indicating that the results are not fully converged with respect to the size of the diagonalized region—the quantum correction remains small in both cases. It amounts to slightly less than 10 \% of the classical Peierls barrier in Fe and less than 2.5 \% in W. We also observe that including additional vibrational modes along the dislocation line in the $2b$ cell does not significantly modify the quantum correction. This indicates that quantum effects on the motion of a straight dislocation across the Peierls barrier are small, in contrast with previous estimates based on EAM potentials \cite{proville2012quantum,barvinschi2014quantum}.
\begin{figure*}[t]
    \centering
   \includegraphics[width=0.95\linewidth]{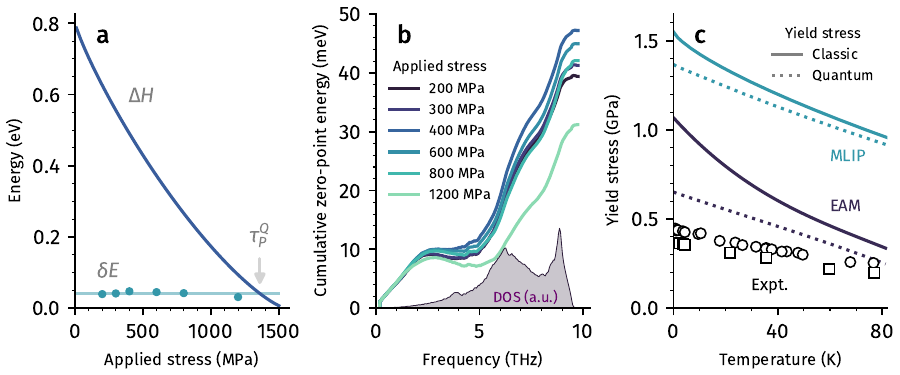}
 \caption{\textbf{Quantum correction to the kink-pair formation enthalpy and flow stress model in Fe.} 
 (a) Kink-pair formation enthalpy ($\Delta H$) and quantum correction ($\delta E$) computed with the MLIP for Fe for different applied stresses.  
 The light horizontal line is the average quantum correction.
 The quantum Peierls stress, at which the barrier effectively vanishes due to zero-point effects, is noted $\tau_\mathrm{P}^\mathrm{Q}$.
 (b) Cumulative quantum correction at different applied stresses. The vibrational density of states of the initial state is reported on the same axis with arbitrary units for easier comparison.
 The final values of the curves correspond to the points shown in (a).
 (c) Flow stress model parametrized using the MLIP and EAM potentials for Fe, compared with experimental results from Ref.~\cite{kubamoto1979thermally}. See Methods section for details.}
\label{fig3}
\end{figure*}

\subsection{MLIP calculations on kinked dislocations}

In order to compute the quantum correction for a long dislocation with a kink-pair profile, representative of the glide mechanism at finite temperature, DFT cannot be used, even with partial Hessian matrices. To access larger simulation cells, we used MLIPs that faithfully reproduce DFT results in small cells. Particular attention was paid to the accuracy of vibrational frequencies that are central in both entropic and quantum zero-point effect. We employed MLIPs for Fe and W recently developed to study {the kink-pair mechanism in thermally activated dislocation glide} \cite{allera2025activation} based on the quadratic formalism introduced in Ref.~\cite{goryaeva2021efficient}. These potentials have already been validated on a number of dislocation properties, including their core structure, Peierls barrier, non-Schmid effects, {and the potential energy landscape of kink-pairs by comparison to a line-tension model parameterized on DFT calculations}. Here, we only discuss vibrational frequencies. The reader is referred to Ref. \cite{allera2025activation} for further information.

Fig.~\ref{fig2} shows the vibrational density of states (DOS) computed in the $1b$ cells containing either the initial or saddle configurations of the dislocation, obtained from  DFT  and from MLIPs using the full Hessian matrices as described above. In the same figure, these DOS are compared to those obtained with two EAM potentials classically used to model dislocations in Fe \cite{proville2012quantum} and W \cite{marinica2013interatomic}. 
As found in Figs. \ref{fig2}(a), (b) , (d) and (e), MLIPs are in excellent agreement with DFT predictions, whereas the EAM potentials exhibit significant deviations. In particular, the EAM potential for Fe predicts a pronounced peak at low frequencies, in contrast to DFT, while the EAM potential for W yields an extended high-frequency tail well beyond the DFT spectrum. To further quantify this comparison, Figs. \ref{fig2}(c) and (f) present the vibrational frequencies obtained with the interatomic potentials (MLIP and EAM) plotted against the corresponding DFT values. We again observe an excellent agreement between DFT and  MLIPs, whereas the EAM results display substantial deviations.

{To assess the effect of such deviations on the quantum correction, we reproduced the $1b$-cell DFT calculation presented in Fig.~1 using the full Hessian matrix, with both the EAM potential and the MLIP in Fe.
The correction is \SI{2.65}{\milli\electronvolt} with the MLIP, \SI{12.30}{\milli\electronvolt} with the EAM potential, and \SI{3.0}{\milli\electronvolt} with DFT, confirming the accuracy of the MLIP.}

Based on their ability to reproduce DFT results,  MLIPs were then used to compute the quantum correction in the \SI{0}{\kelvin} limit for a long dislocation undergoing a transition between Peierls valleys via a kink-pair mechanism at a finite applied shear stress. The initial configuration remains a straight dislocation, while the saddle configuration now consists of a dislocation containing a kink-pair bulge, which breaks the translational symmetry along the line.
We used a parallelepiped simulation cell containing 92,000 atoms. A single screw dislocation was introduced to model a periodic array of dislocations \cite{rodney2007activation,proville2020modeling}, with periodic boundary conditions in the glide plane and free surfaces in the perpendicular direction (see Methods).
The NEB method was again used to relax the minimum energy paths, and the Phondy code (see Methods) was employed to compute the eigenvalues of the full Hessian matrix and the corresponding vibrational frequencies. Also, an external shear stress was applied by imposing uniform forces on the atomic layers at the free surfaces \cite{rodney2007activation,proville2020modeling} to compute $\Delta H(\tau)$ and $\delta E(\tau)$ as a function of the applied stress.

In Fig.~\ref{fig3}(a) we have reported $\Delta H(\tau)$ computed for a $40b$ dislocation with the MLIP for Fe. The enthalpy decreases from about 0.8 eV, the kink-pair formation energy, down to zero at the classical Peierls stress for this potential (about 1.5 GPa). The figure also shows the quantum correction $\delta E$. In contrast to $\Delta H$, $\delta E$ is nearly independent of the applied stress and remains small compared to the kink-pair formation energy (about 0.05 eV). This strongly contrasts with the results obtained using the EAM potential, which predicts a much lower kink-pair formation energy (about \SI{0.54}{\electronvolt}) and yields a much larger, stress-dependent, quantum correction \cite{proville2012quantum}.

We define a quantum Peierls stress $\tau_\mathrm{P}^\mathrm{Q}$ as the stress at which the kink-pair formation enthalpy equals the quantum correction. According to the Fe MLIP, this stress is only about \SI{200}{\mega\pascal} below the classical Peierls stress, corresponding to a reduction of about 13 \%. The predicted quantum Peierls stress thus remains significantly higher than the experimental Peierls stress in Fe (about \SI{450}{\mega\pascal}, \cite{kubamoto1979thermally}). This again contrasts with the EAM potential in Fe, for which the classical Peierls stress is 1000~MPa and the quantum Peierls stress is 650 MPa, much closer to experiments. 
The same calculations were performed with the W MLIP, and confirmed the results obtained in Fe: the quantum correction is small, about \SI{16}{\milli\electronvolt} and independent of the applied stress, leading to a quantum Peierls stress only $5\%$ lower than the classical Peierls stress (see Supplementary Materials for details).

The contribution of the vibrational modes to the overall quantum correction is analyzed in Fig.~\ref{fig3}(b) for Fe and in Supplementary Materials for W, through out the cumulative correction, i.e. the quantum zero point energy variation including modes up to a given frequency. Except at the highest applied stress, the shape of the curves remains remarkably independent of stress, in sharp contrast with the findings of Ref. \cite{proville2012quantum} based on EAM potentials. We also observe that the correction does not originate from specific modes, but rather accumulates over most of the frequency range, with the exception of a plateau between about 2 and \SI{5}{\tera\hertz} and a saturation at the highest frequencies. Furthermore, the low-frequency modes (below \SI{2}{\tera\hertz}), which are presumably extended phonon modes, exhibit a contribution per THz comparable to that of higher-frequency modes, which tend to be more localized (above \SI{5}{\tera\hertz}), as indicated by their similar slopes. However, since the low-frequency modes are fewer, their overall contribution amounts to only about 10 meV, whereas the higher-frequency modes contribute approximately 35 meV.

To further assess the impact of quantum effects on the flow stress in Fe, we incorporate the vibrational frequencies into a quantum formulation of TST, combined with Orowan’s law to predict the temperature dependence of the flow stress for a given applied strain rate. Details of the model are provided in Ref. \cite{proville2012quantum,proville2020modeling} and in the Methods section.
The predicted flow stresses for a plastic shear strain rate of $\dot \epsilon_P = \SI{8e-5}{\per\second}$ are shown in Fig.~\ref{fig3}(c), both with and without quantum effects, and are compared with experimental measurements performed between \SI{0.7}{} and \SI{77}{\kelvin} at the same strain rate~\cite{kubamoto1979thermally}.

We find that both the EAM potential and the MLIP predict, once quantum effects are included, an approximately linear dependence of the flow stress on temperature, in agreement with experiments and in contrast to classical calculations. However, the MLIP predicts flow stresses significantly larger than the experimental values, with only a small quantum correction. In contrast, the EAM potential predicts lower stresses together with a much larger quantum correction, bringing the predicted flow stress close to the experimental data.
These results indicate that according to the current state-of-the-art of dislocation simulations in metals, based on DFT
extended through MLIP,  the 
quantum effects are not the primary physical mechanism responsible for the long-standing discrepancy between atomistic predictions and experiments. Nevertheless, the quantum contribution still amounts to approximately $10\%$ of the classical Peierls stress at very low temperature in Fe and therefore cannot be neglected when aiming for quantitative predictions.

\section{Discussion}

Based on the recent progress in the ability of MLIPs to reproduce the energy landscape of dislocations computed by DFT,
we show that TST-based models, even when corrected with zero-point vibrations, still fail to accurately reproduce the experimental strength of common metals, like iron and tungsten.

Some previous estimates of quantum effects on dislocation glide were performed 
using EAM potentials to model the interatomic forces \cite{proville2012quantum,barvinschi2014quantum,landeiro2017ubiquity,freitas2018quantum,proville2020modeling}, showing
broad variations in the predictions depending on the choice of the EAM model. 
A tentative to account for anharmonic contributions has even been proposed through a ring-polymer method \cite{freitas2018quantum}. Besides its own theoretical limitations regarding real-time quantum dynamics \cite{jang2014can}, the anharmonic contribution does not seem to contribute to enhance quantum effects. 

The main source of uncertainty lies in the accuracy of interatomic potentials, which we rely on to evaluate the quantum correction to the kink-pair nucleation barrier. 
Although MLIPs used in this work substantially improve upon classical EAM descriptions, they remain surrogate models trained on finite DFT datasets, and controlling their extrapolation remains challenging. In particular, invariant descriptors of local atomic environments, such as the bispectrum SO(4) representation \cite{bartok2013representing} used here, may impose intrinsic constraints on the smoothness and topology of the potential energy landscape \cite{dezaphie2025designing}. 
A detailed comparison of crystal defects potential energy landscapes predicted by MLIPs based on different descriptors is an interesting perspective for future work (see e.g. \cite{deng2025systematic}).

It may also be argued that the present MLIPs are sufficiently accurate, such that the remaining limitations originate primarily from the underlying DFT calculations themselves. The DFT methodology employed here corresponds to the current standard for atomistic simulations of dislocations in metals \cite{rodney2017ab,ventelon2013ab,clouet2021screw}. Nevertheless, further improvements may still arise from alternative electronic-structure approximations, including different exchange-correlation functionals. A recent study using meta-GGA functionals for metallic defects reported changes in defect formation energies
\cite{lapointe2025anomalous}.

A third possibility is that the limitations are more fundamental and originate from the theoretical framework itself. The present approach relies on harmonic TST and assumes that quantum effects can be described solely through modifications of crystal vibrational frequencies between stable and saddle configurations. Additional phenomena, non-equilibrium phonon-dislocation coupling, or breakdowns of the harmonic approximation near instability, are not captured within the present framework and could contribute to the remaining discrepancy. Yet, heavy elements such as Fe and W are generally expected to exhibit relatively weak quantum nuclear effects, and the present study confirms that this also remains true for crystal defects and dislocation glide.

Finally, the present study is specifically focused on the migration process of isolated screw dislocations. 
While this approach is corroborated by available in situ transmission microscopy observations \cite{Caillard2010}, the role of other processes, such as collective dislocation interactions and other defect configurations, may warrant further investigation.

Our results suggest that the long-standing discrepancy between atomistic simulations and experimental flow stresses in bcc metals cannot be attributed primarily to zero-point quantum corrections on isolated dislocations forming a kink-pair, contrary to earlier interpretations based on EAM potentials. Rather, they point toward broader limitations of empirical potentials, which appear to poorly reproduce the potential energy landscape of defects in crystals and potentially in other structurally complex systems, such as glasses~\cite{gelin2020enthalpy,li2024infinitely}.

\section*{Methods}

\subsection*{Ab initio calculations}

DFT calculations are performed using the VASP code \cite{kresse1996efficiency}, employing the projector augmented-wave method~\cite{Blochl1994,Kresse1999}. Fe is treated with 8 valence electrons ([Ar] 3d$^7$ 4s$^1$), while W is treated with 6 valence electrons ([Xe 4f$^{14}$] 5d$^5$ 6s$^1$). Spin polarization is included for Fe. Exchange–correlation effects are described within the generalized gradient approximation (GGA) using the Perdew–Burke–Ernzerhof (PBE) functional~\cite{Perdew1996}.
A plane-wave cutoff of 400~eV is used, and Brillouin-zone integrations are performed using the Methfessel-Paxton smearing scheme with a smearing width of 0.2 eV. Atomic structures are relaxed at fixed cell shapes until all components of the ionic forces are below 1~meV/\AA. Minimum energy paths are obtained using the climbing NEB method~\cite{henkelman2000,henkelman2000climbing}, with a spring constant of 5~eV/\AA$^2$, and the atomic positions are relaxed using the same force convergence criterion as in the static calculations.
$2\times2\times16$ and $2\times2\times8$ shifted $k$-point meshes are employed for dislocation line lengths of $1b$ and $2b$, respectively. Hessian matrices are computed via finite differences by displacing atoms by \SI{0.015}{\angstrom} along the $\pm [\bar{1}\bar{1}2]$, $\pm [1\bar{1}0]$, and $\pm [111]$ directions, {which we found satisfactory in our tests, and mass-weighted.
We verified that each transition state is associated with a single imaginary mode.
Translational and imaginary modes are excluded when computing the quantum correction.
The size of the partial Hessian used in Fig.~1 was tested for convergence, as described in Supplementary Materials.}

\subsection*{Classical atomistic simulations}

Classical atomistic simulations are performed using LAMMPS \cite{LAMMPS}. Kink-pair nucleation is modeled in simulation cells containing a periodic array of screw dislocations \cite{bacon2009dislocation}, with a line length $L=40b$ and an in-plane separation of $\sim50b$ along the $\{110\}$ glide plane. The cell height normal to the glide plane is $\sim30b$, with free surfaces in this direction and periodic boundary conditions applied in the glide plane. An external stress is imposed by applying forces to the atomic layers adjacent to the free surfaces \cite{rodney2007activation}. The system contains 92,000 atoms. The same setup was used in Ref. \cite{allera2025activation}. {While this geometry may present finite-size effects in the low stress regime ---where the critical kink-pair is most extended--- the present focus is on moderate to high stress levels, where such effects vanish~\cite{shuang2023effect}.}
Minimum-enthalpy paths for a kinked dislocation are obtained using the NEB method implemented in LAMMPS, with a force tolerance of \SI[per-mode=symbol]{e-3}{\electronvolt\per\angstrom} and a spring constant set to \SI[per-mode=symbol]{1}{\electronvolt\per\angstrom\squared}.
The Hessian matrices and the corresponding eigenvalue spectra are computed using the Phondy code \cite{Marinica2007,proville2012quantum,Soulie2018,Berthier2019,Lapointe2020,Lapointe2022}, employed as an add-on to the LAMMPS molecular dynamics package.
{For the MLIP, the finite displacement magnitude is set to \SI{e-4}{\angstrom}. We verified that using larger displacements did not affect the result.}

\subsection*{Flow stress model}

Under an imposed deformation rate $\dot \epsilon$, the flow stress is the stress value $\tau^*$ that solves Orowan's equation: 
\begin{equation}
\dot \epsilon = \rho b v(\tau^*, T),
\end{equation}
where $\rho$ is the dislocation density and $v$ is the dislocation velocity given by:
\begin{equation}
v = \nu L \exp\left(\frac{-\Delta G(\tau, T)}{kT}\right),
\end{equation}
where $L$ is the dislocation length, $\nu$ is an attempt frequency, $\Delta G$ is the activation free energy for dislocation glide and $k$ is Boltzmann’s constant.
As the present focus is on the low-temperature regime, entropic contributions to the barrier height are neglected, while a stress- and temperature-dependent zero-point energy correction is included. The activation free energy is therefore approximated as $\Delta G (T) \approx \Delta H(\tau) - \delta E(\tau, T)$, where the activation enthalpy follows a Kocks-type form:
\begin{equation}
\Delta H (\tau) = \Delta H_0 \left(1- \left(\frac{\tau}{\tau_P}\right)^p\right)^q,
\end{equation}
as parameterized in our previous work \cite{allera2025activation}.

The stress- and temperature-dependent Wigner correction is expressed following Ref~\cite{proville2012quantum} as:
\begin{align}
    \delta &E(\tau, T) = kT \nonumber\\ &\ln \left( \frac{\prod_{i=1}^{3N-3} 2 \sinh(h\nu_i^{\text{init}}/2kT) / (h\nu_i^{\text{init}}/kT)}{\prod_{k=2}^{3N-4} 2 \sinh(h\nu_k^{*}/2kT) / (h\nu_k^{*}/kT)} \right),
    \label{eq:wigner}
\end{align}
where $\nu^\mathrm{init}$ and $\nu^*$ are the eigenfrequencies at the initial and saddle states, respectively, corresponding to dislocation kink-pair formation under an applied stress $\tau$.
For the MLIP, we found only a weak stress dependence of $\delta E(T)$, and therefore used a value averaged over the different stress levels shown in Fig.~3~(a-b), for simplicity.

The model is applied to the MLIP, using $\Delta H_0 = \SI{0.8}{\electronvolt}$, $\tau_P = \SI{1.55}{\giga\pascal}$, $p = 0.87$, and $q=1.33$.
The deformation rate is set to the experimental value reported in Ref.~\cite{kubamoto1979thermally}, i.e. \SI{8e-5}{\per\second}, the dislocation density is taken as $\rho = \SI{e10}{\per\meter\squared}$ and the prefactor as $\nu = \SI{e15}{\per\second}$ for consistency with Ref.~\cite{proville2012quantum}.
For the EAM potential, a similar flow stress model was previously reported in Ref.~\cite{proville2012quantum}, which we reproduce here in Fig.~3(c) for comparison.

\section*{Data Availability} 
A minimum dataset allowing to reproduce the present calculations, including atomic configurations and interatomic potentials, is available on the Zenodo repository 10.5281/zenodo.15746898~\cite{zenodo}.

\section*{Code availability}
The MILADY package used to perform the MLIP calculations is open-source software, distributed under the Academic Software License (ASL)\cite{milady}. The Phondy code \cite{Marinica2007,  proville2012quantum, Berthier2019, Lapointe2020,Lapointe2022} is part of the MILADY package. 

\section*{Acknowledgments}
Part of this work was performed using HPC resources on Pitagora supercomputer hosted by CINECA.
This work has been carried out within the framework of the EUROfusion Consortium, funded by the European Union via
the Euratom Research and Training Program (Grant Agreement No
101052200-EUROfusion). Views and opinions expressed are however
those of the authors only and do not necessarily reflect those of the
European Union or the European Commission. Neither the European
Union nor the European Commission can be held responsible for them.  
MCM acknowledges the support from GENCI\,-\,(CINES/CCRT) computer centre under Grant No. A0190906973.

\section*{Competing interests}
The authors declare no competing interests.

\bibliography{biblio}

@article{alefeld1964rate,
  title        = {Rate theory in solids at low temperatures},
  author       = {Alefeld, Georg},
  year         = 1964,
  journal      = {Physical Review Letters},
  publisher    = {APS},
  volume       = 12,
  number       = 13,
  pages        = 372
}

@article{allera2025activation,
  title        = {Activation entropy of dislocation glide in body-centered cubic metals from atomistic simulations},
  author       = {Allera, Arnaud and Swinburne, Thomas D and Goryaeva, Alexandra M and Bienvenu, Baptiste and Ribeiro, Fabienne and Perez, Michel and Marinica, Mihai-Cosmin and Rodney, David},
  year         = 2025,
  journal      = {Nature Communications},
  publisher    = {Nature Publishing Group UK London},
  volume       = 16,
  number       = 1,
  pages        = 8367
}

@article{Altshuler1967,
  title        = {The mechanical properties of pure iron tested in compression over the temperature range 2 to 293 K},
  author       = {Altshuler, T. L.  and Christian, J. W.},
  year         = 1967,
  journal      = {Philosophical Transactions of the Royal Society of London. Series A, Mathematical and Physical Sciences},
  volume       = 261,
  number       = 1121,
  pages        = {253--287},
}

@article{aono1981plastic,
  title        = {Plastic deformation of high-purity iron single crystals},
  author       = {Aono, Yasuhisa and Kuramoto, Eiichi and Kitajima, Kazunori},
  year         = 1981,
  journal      = {Reports of Research Institute for Applied Mechanics},
  volume       = 29,
  number       = 92,
  pages        = {127--193}
}

@incollection{bacon2009dislocation,
  title        = {Dislocation–Obstacle Interactions at the Atomic Level},
  author       = {D.J. Bacon and Y.N. Osetsky and D. Rodney},
  year         = 2009,
  booktitle    = {Dislocations in solids},
  publisher    = {Elsevier},
  volume       = 15,
  pages        = {1--90},
  issn         = {1572-4859},
  chapter      = 88,
  editor       = {J.P. Hirth and L. Kubin},
  ldoi         = {https://doi.org/10.1016/S1572-4859(09)01501-0}
}

@article{bartok2013representing,
  title        = {On representing chemical environments},
  author       = {Bart{\'o}k, Albert P and Kondor, Risi and Cs{\'a}nyi, G{\'a}bor},
  year         = 2013,
  journal      = {Physical Review B},
  publisher    = {American Physical Society},
  volume       = 87,
  number       = 18,
  pages        = 184115
}

@article{barvinschi2014quantum,
  title        = {{Quantum Peierls stress of straight and kinked dislocations and effect of non-glide stresses}},
  author       = {Barvinschi, B and Proville, L and Rodney, D},
  year         = 2014,
  journal      = {Modelling and Simulation in Materials Science and Engineering},
  publisher    = {IOP Publishing},
  volume       = 22,
  number       = 2,
  pages        = {025006}
}

@article{basinski1971,
  title        = {Influence of Shear Stress on Screw Dislocations in a Model Sodium Lattice},
  author       = {Basinski, Z. S. and Duesbery, M. S. and Taylor, Roger},
  year         = 1971,
  journal      = {Canadian Journal of Physics},
  volume       = 49,
  number       = 16,
  pages        = {2160--2180}
}

@article{Berthier2019,
  title        = {{Order-disorder or phase-separation transition: Analysis of the Au-Pd system by the effective site energy model}},
  author       = {Berthier, F. and Creuze, J. and Gabard, T. and Legrand, B. and Marinica, M.-C. and Mottet, C.},
  year         = 2019,
  month        = {Jan},
  journal      = {Physical Review B},
  publisher    = {American Physical Society},
  volume       = 99,
  pages        = {014108},
  issue        = 1,
  numpages     = 11
}

@article{bienvenu2022ab,
  title        = {Ab initio informed yield criterion across body-centered cubic transition metals},
  author       = {Bienvenu, Baptiste and Dezerald, Lucile and Rodney, David and Clouet, Emmanuel},
  year         = 2022,
  journal      = {Acta Materialia},
  publisher    = {Elsevier},
  volume       = 236,
  pages        = 118098
}

@article{Blaschke2021,
  title        = {On the temperature and density dependence of dislocation drag from\ phonon wind},
  author       = {Blaschke, Daniel N. and Burakovsky, Leonid and Preston, Dean L.},
  year         = 2021,
  journal      = {Journal of Applied Physics},
  volume       = 130,
  number       = 1,
  pages        = {015901}
}

@article{Blochl1994,
  title        = {{Projector augmented-wave method}},
  author       = {Bl{\"{o}}chl, P E},
  year         = 1994,
  journal      = {Physical Review B},
  volume       = 50,
  number       = 24,
  pages        = {17953--17979}
}

@article{brunner2000plastic,
  title        = {{The plastic properties of high-purity W single crystals}},
  author       = {Brunner, D and Glebovsky, V},
  year         = 2000,
  journal      = {Materials Letters},
  publisher    = {Elsevier},
  volume       = 42,
  number       = 5,
  pages        = {290--296}
}

@book{caillard2003thermally,
  title        = {Thermally activated mechanisms in crystal plasticity},
  title        = {Thermally Activated Mechanisms in Crystal Plasticity},
  author       = {Caillard, Daniel and Martin, Jean-Luc},
  year         = 2003,
  publisher    = {Pergamon Press},
  series       = {Pergamon Materials Series},
  volume       = 8
}

@article{Caillard2010,
  title        = {{Kinetics of dislocations in pure Fe. Part II. In situ straining experiments at low temperature}},
  author       = {Caillard, D},
  year         = 2010,
  journal      = {Acta Materialia},
  publisher    = {Elsevier},
  volume       = 58,
  number       = 9,
  pages        = {3504--3515}
}

@article{caillard2018geometry,
  title        = {{Geometry and kinetics of glide of screw dislocations in tungsten between 95~K and 573~K}},
  author       = {Caillard, Daniel},
  year         = 2018,
  journal      = {Acta Materialia},
  publisher    = {Elsevier},
  volume       = 161,
  pages        = {21--34}
}

@article{cereceda2016unraveling,
  title        = {Unraveling the temperature dependence of the yield strength in single-crystal tungsten using atomistically-informed crystal plasticity calculations},
  author       = {Cereceda, David and Diehl, Martin and Roters, Franz and Raabe, Dierk and Perlado, J Manuel and Marian, Jaime},
  year         = 2016,
  journal      = {International Journal of Plasticity},
  publisher    = {Elsevier},
  volume       = 78,
  pages        = {242--265}
}

@article{chaussidon2006glide,
  title        = {{The glide of screw dislocations in bcc Fe: atomistic static and dynamic simulations}},
  author       = {Chaussidon, Julien and Fivel, Marc and Rodney, David},
  year         = 2006,
  journal      = {Acta Materialia},
  publisher    = {Elsevier},
  volume       = 54,
  number       = 13,
  pages        = {3407--3416}
}

@article{clouet2021screw,
  title        = {{Screw dislocations in BCC transition metals: from ab initio modeling to yield criterion}},
  author       = {Clouet, Emmanuel and Bienvenu, Baptiste and Dezerald, Lucile and Rodney, David},
  year         = 2021,
  journal      = {Comptes Rendus Physique},
  volume       = 22,
  number       = {S3},
  pages        = {83--116}
}

@article{deng2025systematic,
  title        = {Systematic softening in universal machine learning interatomic potentials},
  author       = {Deng, Bowen and Choi, Yunyeong and Zhong, Peichen and Riebesell, Janosh and Anand, Shashwat and Li, Zhuohan and Jun, KyuJung and Persson, Kristin A and Ceder, Gerbrand},
  year         = 2025,
  journal      = {npj Computational Materials},
  publisher    = {Nature Publishing Group},
  volume       = 11,
  number       = 1,
  pages        = {1--9}
}

@article{dezaphie2025designing,
  title        = {Designing hybrid descriptors for improved machine learning models in atomistic materials science simulations},
  author       = {Dezaphie, Alexandre and Lapointe, Clovis and Goryaeva, Alexandra M and Creuze, J{\'e}r{\^o}me and Marinica, Mihai-Cosmin},
  year         = 2025,
  journal      = {Computational Materials Science},
  publisher    = {Elsevier},
  volume       = 246,
  pages        = 113459
}

@article{dezerald2016plastic,
  title        = {{Plastic anisotropy and dislocation trajectory in BCC metals}},
  author       = {Dezerald, Lucile and Rodney, David and Clouet, Emmanuel and Ventelon, Lisa and Willaime, Francois},
  year         = 2016,
  journal      = {Nature Communications},
  publisher    = {Nature Publishing Group},
  volume       = 7,
  number       = 1,
  pages        = {1--7}
}

@article{freitas2018quantum,
  title        = {Quantum effects on dislocation motion from ring-polymer molecular dynamics},
  author       = {Freitas, Rodrigo and Asta, Mark and Bulatov, Vasily V},
  year         = 2018,
  journal      = {npj Computational Materials},
  publisher    = {Nature Publishing Group UK London},
  volume       = 4,
  number       = 1,
  pages        = 55
}

@article{gelin2020enthalpy,
  title        = {{Enthalpy-entropy compensation of atomic diffusion originates from softening of low frequency phonons}},
  author       = {Gelin, Simon and Champagne-Ruel, Alexandre and Mousseau, Normand},
  year         = 2020,
  journal      = {Nature Communications},
  publisher    = {Nature Publishing Group},
  volume       = 11,
  number       = 1,
  pages        = 3977
}

@article{glen1956xxxix,
  title        = {XXXIX. The creep of cadmium crystals at liquid helium temperatures},
  author       = {Glen, JW},
  year         = 1956,
  journal      = {Philosophical Magazine},
  publisher    = {Taylor \& Francis},
  volume       = 1,
  number       = 5,
  pages        = {400--408}
}

@article{goryaeva2021efficient,
  title        = {{Efficient and transferable machine learning potentials for the simulation of crystal defects in bcc Fe and W}},
  author       = {Goryaeva, Alexandra M and D{\'e}r{\`e}s, Julien and Lapointe, Clovis and Grigorev, Petr and Swinburne, Thomas D and Kermode, James R and Ventelon, Lisa and Baima, Jacopo and Marinica, Mihai-Cosmin},
  year         = 2021,
  journal      = {Physical Review Materials},
  publisher    = {American Physical Society},
  volume       = 5,
  number       = 10,
  pages        = 103803
}

@article{groger2007explanation,
  title        = {Explanation of the discrepancy between the measured and atomistically calculated yield stresses in body-centred cubic metals},
  author       = {Gr{\"o}ger, R and Vitek, V},
  year         = 2007,
  journal      = {Philosophical Magazine Letters},
  publisher    = {Taylor \& Francis},
  volume       = 87,
  number       = 2,
  pages        = {113--120}
}

@article{guyot1967critical,
  title        = {{A critical review of the Peierls mechanism}},
  author       = {Guyot, Pierre and Dorn, John E},
  year         = 1967,
  journal      = {Canadian Journal of Physics},
  publisher    = {NRC Research Press Ottawa, Canada},
  volume       = 45,
  number       = 2,
  pages        = {983--1016}
}

@article{henkelman2000,
  title        = {{Improved tangent estimate in the nudged elastic band method for finding minimum energy paths and saddle points}},
  author       = {Henkelman, G and Jonsson, H},
  year         = {{2000}},
  journal      = {{The Journal of Chemical Physics}},
  volume       = {{113}},
  number       = {{22}},
  pages        = {{9978--9985}}
}

@article{henkelman2000climbing,
  title        = {{A climbing image nudged elastic band method for finding saddle points and minimum energy paths}},
  author       = {Henkelman, Graeme and Uberuaga, Blas P and J{\'o}nsson, Hannes},
  year         = 2000,
  journal      = {The Journal of Chemical Physics},
  publisher    = {American Institute of Physics},
  volume       = 113,
  number       = 22,
  pages        = {9901--9904}
}

@article{jang2014can,
  title        = {Can the ring polymer molecular dynamics method be interpreted as real time quantum dynamics?},
  author       = {Jang, Seogjoo and Sinitskiy, Anton V and Voth, Gregory A},
  year         = 2014,
  journal      = {The Journal of Chemical Physics},
  publisher    = {AIP Publishing},
  volume       = 140,
  number       = 15
}

@article{kamimura2013experimental,
  title        = {Experimental evaluation of the Peierls stresses in a variety of crystals and their relation to the crystal structure},
  author       = {Kamimura, Y and Edagawa, K and Takeuchi, S},
  year         = 2013,
  journal      = {Acta materialia},
  publisher    = {Elsevier},
  volume       = 61,
  number       = 1,
  pages        = {294--309}
}

@article{kamimura2018peierls,
  title        = {{Peierls stresses estimated via the Peierls-Nabarro model using ab-initio $\gamma$-surface and their comparison with experiments}},
  author       = {Kamimura, Y and Edagawa, K and Iskandarov, AM and Osawa, M and Umeno, Y and Takeuchi, S},
  year         = 2018,
  journal      = {Acta Materialia},
  publisher    = {Elsevier},
  volume       = 148,
  pages        = {355--362}
}

@article{kang2012singular,
  title        = {Singular orientations and faceted motion of dislocations in body-centered cubic crystals},
  author       = {Kang, Keonwook and Bulatov, Vasily V and Cai, Wei},
  year         = 2012,
  journal      = {Proceedings of the National Academy of Sciences},
  publisher    = {National Academy of Sciences},
  volume       = 109,
  number       = 38,
  pages        = {15174--15178}
}

@article{kresse1996efficiency,
  title        = {Efficiency of ab-initio total energy calculations for metals and semiconductors using a plane-wave basis set},
  author       = {Kresse, Georg and Furthm{\"u}ller, J{\"u}rgen},
  year         = 1996,
  journal      = {Computational Materials Science},
  publisher    = {Elsevier},
  volume       = 6,
  number       = 1,
  pages        = {15--50}
}

@article{Kresse1999,
  title        = {From ultrasoft pseudopotentials to the projector augmented-wave method},
  author       = {Kresse, G. and Joubert, D.},
  year         = 1999,
  journal      = {Physical Review B},
  volume       = 59,
  number       = 3,
  pages        = {1758--1775}
}

@article{kubamoto1979thermally,
  title        = {Thermally activated slip deformation between 0.7 and 77 {K} in high-purity iron single crystals},
  author       = {Kubamoto, E and Aono, Y and Kitajima, K and Maeda, K and Takeuchi, S},
  year         = 1979,
  journal      = {Philosophical Magazine A},
  publisher    = {Taylor \& Francis},
  volume       = 39,
  number       = 6,
  pages        = {717--724}
}

@article{LAMMPS,
  title        = "{LAMMPS} - a flexible simulation tool for
     particle-based materials modeling at the 
     atomic, meso, and continuum scales",
  author       = "A. P. Thompson and H. M. Aktulga and R. Berger and 
     D. S. Bolintineanu and W. M. Brown and P. S. Crozier and
     P. J. in 't Veld and A. Kohlmeyer and S. G. Moore and T. D. Nguyen and
     R. Shan and M. J. Stevens and J. Tranchida and C. Trott and S. J. Plimpton",
  year         = 2022,
  journal      = "Computer Physics Communications",
  volume       = 271,
  pages        = 108171,
  ldoi         = "10.1016/j.cpc.2021.108171"
}

@article{landeiro2017ubiquity,
  title        = {Ubiquity of quantum zero-point fluctuations in dislocation glide},
  author       = {Landeiro Dos Reis, Marie and Choudhury, Anshuman and Proville, Laurent},
  year         = 2017,
  month        = {Mar},
  journal      = {Phys. Rev. B},
  publisher    = {American Physical Society},
  volume       = 95,
  pages        = {094103},
  issue        = 9,
  numpages     = 8
}

@article{Lapointe2020,
  title        = {Machine learning surrogate models for prediction of point defect vibrational entropy},
  author       = {Lapointe, Clovis and Swinburne, Thomas D. and Thiry, Louis and Mallat, St\'ephane and Proville, Laurent and Becquart, Charlotte S. and Marinica, Mihai-Cosmin},
  year         = 2020,
  month        = {Jun},
  journal      = {Physical Review Materials},
  publisher    = {American Physical Society},
  volume       = 4,
  pages        = {063802},
  issue        = 6,
  numpages     = 12
}

@article{Lapointe2022,
  title        = {Machine learning surrogate models for strain-dependent vibrational properties and migration rates of point defects},
  author       = {Lapointe, Clovis and Swinburne, Thomas D. and Proville, Laurent and Becquart, Charlotte S. and Mousseau, Normand and Marinica, Mihai-Cosmin},
  year         = 2022,
  month        = {Nov},
  journal      = {Physical Review Materials},
  publisher    = {American Physical Society},
  volume       = 6,
  pages        = 113803,
  issue        = 11,
  numpages     = 15
}

@article{lapointe2025anomalous,
  title        = {Anomalous self-diffusion in tungsten and molybdenum: Exonerating the di-vacancy contribution and the key role of interatomic interaction},
  author       = {Lapointe, Clovis and Zhong, Anruo and Swinburne, Thomas D and Bruneval, Fabien and Ath{\`e}nes, Manuel and Marinica, Mihai-Cosmin},
  year         = 2025,
  journal      = {Physical Review Materials},
  publisher    = {American Physical Society},
  volume       = 9,
  number       = 9,
  pages        = {093801}
}

@article{lerma2025quantum,
  title        = {Quantum statistical theory of dislocation mobility in discrete lattices},
  author       = {Lerma, Be{\~n}at Gurrutxaga},
  year         = 2025,
  journal      = {Physical Review Materials},
  publisher    = {American Physical Society},
  volume       = 9,
  number       = 12,
  pages        = 123605
}

@article{li2017nonperturbative,
  title        = {Nonperturbative quantum nature of the dislocation-phonon interaction},
  author       = {Li, Mingda and Ding, Zhiwei and Meng, Qingping and Zhou, Jiawei and Zhu, Yimei and Liu, Hong and Dresselhaus, Mildred S and Chen, Gang},
  year         = 2017,
  journal      = {Nano Letters},
  publisher    = {ACS Publications},
  volume       = 17,
  number       = 3,
  pages        = {1587--1594}
}

@article{li2024infinitely,
  title        = {Infinitely rugged intra-cage potential energy landscape in metallic glasses caused by many-body interaction},
  author       = {Li, Haoyu and Xiao, Hongyi and Egami, Takeshi and Fan, Yue},
  year         = 2024,
  journal      = {Materials Today Physics},
  publisher    = {Elsevier},
  volume       = 49,
  pages        = 101582
}

@article{maresca2018screw,
  title        = {{Screw dislocation structure and mobility in body centered cubic Fe predicted by a Gaussian Approximation Potential}},
  author       = {Maresca, Francesco and Dragoni, Daniele and Cs{\'a}nyi, G{\'a}bor and Marzari, Nicola and Curtin, William A},
  year         = 2018,
  journal      = {npj Computational Materials},
  publisher    = {Nature Publishing Group UK London},
  volume       = 4,
  number       = 1,
  pages        = 69
}

@article{Marinica2007,
  title        = {{Orientation of interstitials in clusters in $\alpha$-Fe: a comparison between empirical potentials}},
  author       = {Mihai Cosmin Marinica and F. Willaime},
  year         = 2007,
  journal      = {Solid State Phenomena},
  publisher    = {Trans Tech Publications},
  volume       = 129,
  pages        = {67--74}
}

@article{marinica2013interatomic,
  title        = {{Interatomic potentials for modelling radiation defects and dislocations in tungsten}},
  author       = {Marinica, Mihai-Cosmin and Ventelon, Lisa and Gilbert, MR and Proville, L and Dudarev, SL and Marian, J and Bencteux, G and Willaime, F},
  year         = 2013,
  journal      = {Journal of Physics: Condensed Matter},
  publisher    = {IOP Publishing},
  volume       = 25,
  number       = 39,
  pages        = 395502
}

@misc{milady,
  title        = {},
  author       = {},
  year         = {},
  note         = " ",
  howpublished = "\url{https://ai-atoms.github.io/milady/}"
}

@article{Mott1956-MOTLCI,
  title        = {Lvii. Creep in Metal Crystals at Very Low Temperatures},
  author       = {N. F. Mott},
  year         = 1956,
  journal      = {Philosophical Magazine},
  volume       = 1,
  number       = 6,
  pages        = {568--572}
}

@article{natsik1972influence,
  title        = {The Influence of Quantum Effects on the Low-Temperature Creep of Zinc Crystals},
  author       = {Natsik, VD and Osetskii, AI and Soldatov, VP and Startsev, VI},
  year         = 1972,
  journal      = {physica status solidi (b)},
  publisher    = {Wiley Online Library},
  volume       = 54,
  number       = 1,
  pages        = {99--111}
}

@article{Perdew1996,
  title        = {Generalized Gradient Approximation Made Simple},
  author       = {John P. Perdew and Kieron Burke and Matthias Ernzerhof},
  year         = 1996,
  month        = {oct},
  journal      = {Physical Review Letters},
  volume       = 77,
  number       = 18,
  pages        = {3865--3868}
}

@article{proville2012quantum,
  title        = {{Quantum effect on thermally activated glide of dislocations}},
  author       = {Proville, Laurent and Rodney, David and Marinica, Mihai-Cosmin},
  year         = 2012,
  journal      = {Nature Materials},
  publisher    = {Nature Publishing Group},
  volume       = 11,
  number       = 10,
  pages        = {845--849}
}

@inbook{proville2020modeling,
  title        = {Modeling the Thermally Activated Mobility of Dislocations at the Atomic Scale},
  author       = {Proville, Laurent and Rodney, David},
  year         = 2020,
  publisher    = {Springer International Publishing},
  address      = {Cham},
  pages        = {1525--1544},
  editor       = {Andreoni, Wanda and Yip, Sidney}
}

@article{rodney2007activation,
  title        = {{Activation enthalpy for kink-pair nucleation on dislocations: Comparison between static and dynamic atomic-scale simulations}},
  author       = {Rodney, David},
  year         = 2007,
  journal      = {Physical Review B},
  publisher    = {American Physical Society},
  volume       = 76,
  number       = 14,
  pages        = 144108
}

@article{rodney2017ab,
  title        = {Ab initio modeling of dislocation core properties in metals and semiconductors},
  author       = {Rodney, David and Ventelon, L and Clouet, E and Pizzagalli, Laurent and Willaime, F},
  year         = 2017,
  journal      = {Acta Materialia},
  publisher    = {Elsevier},
  volume       = 124,
  pages        = {633--659}
}

@article{seeger2022peierls,
  title        = {{Peierls barriers, kinks, and flow stress: Recent progress}},
  author       = {Seeger, Alfred},
  year         = 2022,
  journal      = {International Journal of Materials Research},
  publisher    = {De Gruyter},
  volume       = 93,
  number       = 8,
  pages        = {760--777}
}

@article{Soulie2018,
  title        = {{Influence of vibrational entropy on the concentrations of oxygen interstitial clusters and uranium vacancies in nonstoichiometric $\mathrm{U}{\mathrm{O}}_{2}$}},
  author       = {Souli\'e, Aur\'elien and Bruneval, Fabien and Marinica, Mihai-Cosmin and Murphy, Samuel and Crocombette, Jean-Paul},
  year         = 2018,
  month        = {Aug},
  journal      = {Physical Review Materials},
  publisher    = {American Physical Society},
  volume       = 2,
  pages        = {083607},
  issue        = 8,
  numpages     = 7
}

@inproceedings{Suzuki1970Effect,
  title        = {Effect of Zero-Point Motion on Peierls Stress},
  author       = {Suzuki, H.},
  year         = 1970,
  booktitle    = {Fundamental Aspects of Dislocation Theory},
  publisher    = {U.S. Government Printing Office, Washington, D.C.},
  series       = {National Bureau of Standards Special Publication},
  number       = {317, Volume 1},
  pages        = {253--272},
  note         = {Conference held April 21--25, 1969},
  editor       = {Simmons, John A. and deWit, Roland and Bullough, R.}
}

@article{tallman2019hierarchical,
  title        = {{Hierarchical top-down bottom-up calibration with consideration for uncertainty and inter-scale discrepancy of Peierls stress of bcc Fe}},
  author       = {Tallman, Aaron E and Swiler, Laura P and Wang, Yan and McDowell, David L},
  year         = 2019,
  journal      = {Modelling and Simulation in Materials Science and Engineering},
  publisher    = {IOP Publishing},
  volume       = 27,
  number       = 6,
  pages        = {064004}
}

@article{ventelon2007core,
  title        = {{Core structure and Peierls potential of screw dislocations in $\alpha$-{F}e from first principles: cluster versus dipole approaches}},
  author       = {Ventelon, Lisa and Willaime, F},
  year         = 2007,
  journal      = {Journal of Computer-Aided Materials Design},
  publisher    = {Springer},
  volume       = 14,
  pages        = {85--94}
}

@article{ventelon2013ab,
  title        = {{Ab initio investigation of the Peierls potential of screw dislocations in bcc Fe and W}},
  author       = {Ventelon, Lisa and Willaime, Fran{\c{c}}ois and Clouet, Emmanuel and Rodney, David},
  year         = 2013,
  journal      = {Acta Materialia},
  publisher    = {Elsevier},
  volume       = 61,
  number       = 11,
  pages        = {3973--3985}
}

@article{weinberger2013peierls,
  title        = {{Peierls potential of screw dislocations in bcc transition metals: Predictions from density functional theory}},
  author       = {Weinberger, Christopher R and Tucker, Garritt J and Foiles, Stephen M},
  year         = 2013,
  journal      = {Physical Review B},
  publisher    = {American Physical Society},
  volume       = 87,
  number       = 5,
  pages        = {054114}
}

@article{weygand2015multiscale,
  title        = {Multiscale simulation of plasticity in bcc metals},
  author       = {Weygand, Daniel and Mrovec, Matous and Hochrainer, Thomas and Gumbsch, Peter},
  year         = 2015,
  journal      = {Annual Review of Materials Research},
  publisher    = {Annual Reviews},
  volume       = 45,
  number       = 1,
  pages        = {369--390}
}

@misc{zenodo,
  title        = {Supporting files for "Activation entropy of dislocation glide in body-centered cubic metals from atomistic simulations"},
  author       = {Allera, Arnaud and Swinburne, Thomas and Goryaeva, Alexandra M. and Bienvenu, Baptiste and Ribeiro, Fabienne and Perez, Michel and Marinica, Mihai-Cosmin and Rodney, David},
  year         = 2025,
  note         = "",
  howpublished = "10.5281/zenodo.15746898"
}

@article{ZHAO2024105640,
  title        = {Multiscale modeling of dislocation-mediated plasticity of refractory high entropy alloys},
  author       = {Feng Zhao and Wenbin Liu and Xin Yi and Yin Zhang and Huiling Duan},
  year         = 2024,
  journal      = {Journal of the Mechanics and Physics of Solids},
  volume       = 187,
  pages        = 105640,
  issn         = {0022-5096},
  ldoi         = {https://doi.org/10.1016/j.jmps.2024.105640},
  lurl         = {https://www.sciencedirect.com/science/article/pii/S0022509624001066}
}

@article{shuang2023effect,
  title={Effect of periodic image interactions on kink pair activation of screw dislocation},
  author={Shuang, Fei and Ji, Rigelesaiyin and Xiong, Liming and Gao, Wei},
  journal={Computational Materials Science},
  volume={228},
  pages={112369},
  year={2023},
  publisher={Elsevier}
}
\end{document}


\begin{frontmatter}

\title{
Supplementary Materials: 
Revisiting quantum effects on dislocation glide in bcc metals from DFT calculations and machine-learning potentials
}

\author[label1]{Arnaud Allera\corref{cor1}}
\ead{arnaud.allera@asnr.fr}
\author[label3]{Lisa Ventelon}
\ead{lisa.ventelon@cea.fr}
\author[label3]{Mihai-Cosmin Marinica}
\ead{mihai-cosmin.marinica@cea.fr}
\author[label2]{David Rodney}
\ead{david.rodney@univ-lyon1.fr}
\author[label3]{Laurent Proville}
\ead{laurent.proville@cea.fr}

\address[label1]{ASNR/PSN-RES/SEMIA/LSMA Centre d'études de Cadarache, F-13115 Saint Paul-lez-Durance, France}
\address[label3]{Université Paris-Saclay, CEA, Service de recherche en Corrosion et Comportement des Matériaux, SRMP, 91191 Gif Sur Yvette, France}
\address[label2]{Univ. Lyon, UCBL, Institut Lumière Matière, UMR CNRS 5306, F-69622 Villeurbanne, France}

\end{frontmatter}

\setcounter{figure}{0}
\renewcommand{\thefigure}{S\arabic{figure}}
\renewcommand{\figurename}{Supplementary Figure}

\section{Complementary results in W}

\begin{figure}[hb]
    \centering
    \includegraphics[width=0.7\linewidth]{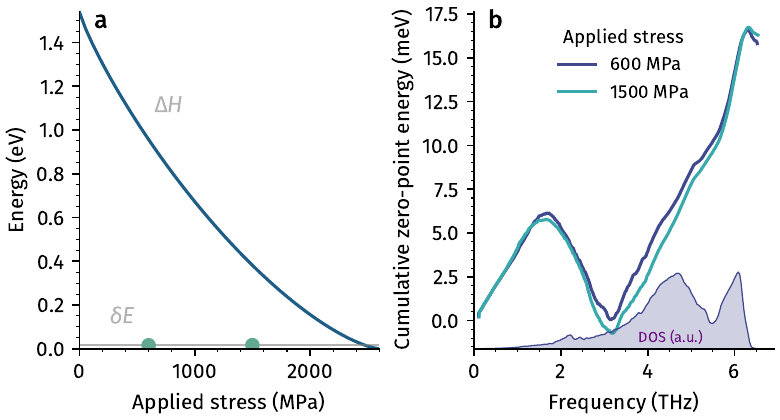}
    \caption{\textbf{Quantum correction to the kink-pair formation enthalpy in W.} 
     (a) Kink-pair formation enthalpy ($\Delta H$) and quantum correction ($\delta E$) for W for different applied stresses.  
     The light horizontal line represents the average quantum correction. (b) Cumulative quantum correction at different applied stresses.
     }
    \label{fig:supp_PAD_W}
\end{figure}

The results presented in the main text clearly indicate a minor ZPE correction for dislocation glide in Fe, both on the Peierls barrier and kink-pair nucleation.
To explore the generality of our main result,
we reproduce the kink-pair nucleation calculations in W, using a MLIP recently developed for accurate dislocation simulations \cite{allera2025activation}.
As seen on Fig.~\ref{fig:supp_PAD_W} (a), the ZPE correction is approximately \SI{16}{\milli\electronvolt} at 600 and \SI{1500}{\mega\pascal}, representing a minor reduction of the Peierls stress by about $5\%$.
The correction is significantly lower than in Fe, which is consistent with direct DFT calculations on straight dislocation segments (see Figs.~1(b) and (c) of the main text).
For completeness, Fig.~\ref{fig:supp_PAD_W} (b) presents the cumulated ZPE difference between the initial and saddle state, showing a remarkingly stress-independent behavior, again in stark contrast with reports based on EAM potentials.
The profile of the cumulated $\delta E(\nu)$ with respect to frequency is mostly similar to the results in Fe (Fig.~3~(b) of the main text), with a more pronounced increase near $\max(\nu)/2$ and less important variations, resulting in a lower correction $\delta E$.

\section{Convergence of partial Hessian matrix calculations}

We computed the vibration modes of a crystal with a migrating screw dislocation, using direct DFT calculations.
The computational cost of this operation increases rapidly with the number of atoms considered in the calculation.
To keep DFT calculations tractable, especially for the $2b$-thick simulation cells presented in the main text, we computed the ZPE using a partial Hessian matrix, where only a subset of the atoms is displaced to construct the dynamical matrix via finite differences \cite{barvinschi2014quantum}.

To assess the convergence rate of this approach, we performed a series of partial Hessian calculations using the Fe MLIP in a 135-atoms, $1b$-thick cell containing a dislocation dipole. 
We included an increasingly large number of atoms surrounding one dislocation core, and compared the resulting ZPE difference between the initial and saddle state to a full calculation including all atoms. For each setting, the same set of atoms is displaced at both the initial and saddle states.
The minimal set of atoms we consider is restricted to the core atoms, i.e. it includes the atoms closest to the dislocation core in its initial state (3 atoms) and at the saddle state (3 atoms, 2 of which are shared with the initial state), resulting in a set of 4 atoms (in red on Fig. S2 a)).
Additional atomic layers surrounding this minimal set are then successively included---denoted $n$-th nearest neighbors (NN) in Fig. S2---up to third-nearest neighbors.
In Fig. S2 a), we show the first NN layer atoms in gray. 

As shown in Fig. S2 b), we found that including the first nearest neighbors of the core atoms reduces the error threefold compared to only considering the core atoms, reducing the error to \SI{1}{\milli\electronvolt}, which is much smaller than the Peierls barrier. 
Beyond this point, the error decreases only logarithmically with the number of atomic layers included. We conclude that including atoms up to the first NN shell (yielding 14 atoms) provides the best compromise between computational cost and accuracy. We therefore adopt this setting for the partial Hessian calculations presented in the main text.

\begin{figure}[h]
    \centering
    
    \includegraphics[width=0.39\linewidth]{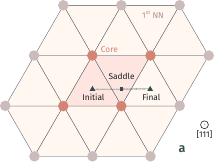}\hfill
    \includegraphics[width=0.54\linewidth]{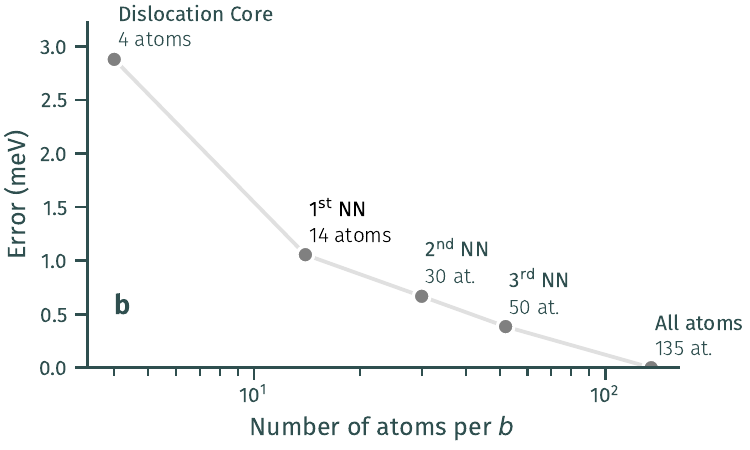}
    \caption{\textbf{Partial Hessian matrix size optimization.} 
    a) Schematic representation of the atomic positions in the (111) plane, showing the position of the dislocation core along the Peierls barrier at the initial, saddle and final states (triangles and square symbols). 
    The atoms closest to the dislocation core at the initial position or at the saddle state are shown in red and denoted as "core" atoms. 
    The atoms belonging to the first atomic layer around core atoms are denoted as "1$^\mathrm{st}$ NN" (in gray).
    b) The error on the ZPE evaluted using a partial Hessian relative to a full calculation, $|\delta E_\mathrm{full} - \delta E_\mathrm{partial}|$, is plotted against the number of atoms, which drives the computational cost.
    }
    \label{fig:placeholder}
\end{figure}

\bibliography{biblio}